\documentstyle[12pt]{article}
\begin{document}
\begin {center}
{\bf {\Large  $\pi^+\pi^-$ Emission from Coherent $\rho^0$ Propagation
in Nuclei} }
\end {center}
\begin {center}
Swapan Das  \\
{\it Nuclear Physics Division,
Bhabha Atomic Research Centre  \\
Mumbai-400085, India }
\end {center}

\begin {abstract}
Cross sections for the $\pi^+\pi^-$ events arising due to the decay of
$\rho^0$ meson have been calculated. This rho meson is considered to produce
coherently in the $(p,p^\prime)$ reaction on a nucleus.
The distorted wave functions for the continuum particles and the rho meson
propagator are described by the eikonal form.
The sensitivity of the cross section on the pion emission angle as well as
on the beam energy has been investigated.
The cross sections for the $\rho^0$ decaying inside and outside the nucleus
are compared. In addition, the coherent and incoherent contributions to the
cross section due to $\rho^0$ inside and outside decay amplitudes are
presented.
The initial and final state interactions for this reaction have been studied. The
nuclear medium effect on the $\rho^0$ propagation through the nucleus is
also explored.
\end {abstract}

\section {Introduction}
\label {Int}

~~~~
The hadron parameters for vector mesons embedded in the nucleus may differ
significantly from their free space values due to the interaction taken
place between the vector mesons and the other nuclear particles present
in the nucleus. Therefore, the study on the in-medium properties of vector
mesons can offer valuable informations about the vector meson dynamics in a
nucleus. In compressed [$\varrho \approx (3-5) \varrho_0 $] and/or hot
[$\sim ~ (150-200)$ MeV] nuclei, the medium effect on the vector meson could be drastic.
In fact, large medium effect on the rho meson is believed to be seen in the
enhanced dilepton yield in CERES and HELIOS ultra-relativistic heavy ion
collision data \cite{dilpex} taken in CERN-SPS.
Theoretically, these data are found to be compatible if the $\rho$ mass
is reduced drastically, i.e., by $\sim ~ 300-400$ MeV \cite{dfms}. In
contrast to this, these data are also reproduced successfully by calculations
\cite{mbcl} which incorporate the hadronic interaction for the rho meson with
the surrounding nuclear particles. Recent analysis on the $\pi^+\pi^-$ pairs,
in the STAR experiment at BNL RHIC, showed a decrease in $\rho$ mass in
the peripheral Au + Au collisions \cite{rhic}.

There exist some model calculations which envisage the reduction of 
vector meson mass in the nuclear medium. For example, the scaling hypothesis
due to Brown and Rho \cite{br} shows that the mass of vector meson in a
nucleus should drop, since the pion weak decay constant does. The QCD sum
rule calculation done by Hatsuda and Lee \cite{hat} shows that the reduction
in the vector meson mass increases with the nuclear density. This finding is
also corroborated by the vector dominance model (VDM) calculation due to
Asakawa et al., \cite{asa}. The quark meson (QM) coupling model calculation
done by Saito et al., \cite{sai} shows that the mass of $\rho$ meson in the
$^{12}$C nucleus is reduced by 50 MeV.
Beside these, the manybody calculations show that the spectral function for
the rho meson in the nuclear medium gets modified significantly due to its
interaction with the nucleon and resonances.
In one of these calculations, Frimann et al., \cite{fpr} have shown that the
$p$ wave $\rho N$ scattering [via $N(1720)$ and $\Delta (1905)$ resonances]
reduces the mass of rho meson, and this reduction is significant in high
baryon density.
Peters et al., \cite{ptr} have extended this calculations by incorporating
all four starred $s$ and $p$ wave resonances. Their calculations show a
strong influence of the $s$ wave resonances on the $\rho$ spectral function,
specifically in the static limit.
The density dependent two pion self-energy calculation due to Herrmann et
al., \cite{her} shows an enhancement in the in-medium $\rho$ width, but they
have not found, unlike others, any significant change in the in-medium $\rho$
mass. The recent developments in this field are illustrated in the
Ref.~\cite{redv}.

To look for the in-medium properties of vector mesons in a normal nucleus,
experimental programmes are there at various centers around the world.
For example, it has been proposed to measure the photo production of
vector meson in the nucleus at TJNAF through the $e^+e^-$ radiation
\cite{tjn}, and at 1.3 GeV Tokyo Electron Synchrotron (INS) through the
$\pi^+\pi^-$ emission \cite{ins}.
Recently, CBELSA/TAPS Collaboration at ELSA \cite{elsa} had found the
medium modification on the omega meson in $^{93}$Nb nucleus.
The KEK-PS E325 collaboration at KEK \cite{epem} had measured the $e^+e^-$
yield in the $p+A$ reaction at 12 GeV, and they have reported an enhancement
in this yield in the region of $ 0.6 \ge m_{e^+e^-} \le 0.77 $ GeV/c$^2$
due to modified $\rho$ meson in nuclei.
Another experiment \cite{phi} in this collaboration (which investigates the
medium effect on $\phi$ meson) is in progress.
There also exist experimental programmes at GSI-SIS \cite{hade} (HADE
collaboration) as well as at Spring-8 RCNP \cite{sprn}, which are dedicated
to explore the hadron parameters for vector mesons in the normal nucleus.
Therefore, the overall view for the medium modification on vector mesons is
expected in near future.

There also exist several calculations showing the medium effect on vector
mesons in the normal nucleus. Eletsky et al., \cite{elio} and Kondratyuk et
al., \cite{kscge} relate the effective mass and width for the rho meson in a
nucleus to the $\rho$ nucleon scattering amplitude $f_{\rho N}$ and nuclear
density.
In their calculations, it has been shown that the mass of the rho meson in
a normal nucleus increases with its energy. In the static limit, the $\rho$
mass is below its free space value, where as it exceeds 770 MeV at higher
energies \cite{kscge}.
Effenberger et al., have studied the $e^+e^-$ production from the gamma
\cite{effn1} and pion \cite{effn2} induced nuclear reactions at GeV energies
in the framework of semiclassical Boltzmann-Uehling-Uhlenbeck (BUU) transport
model. Their study shows that the medium modification on vector mesons
enhances the dilepton yields in the $e^+e^-$ invariant mass region below 770
MeV.
The Intranuclear Cascade (INC) model calculation due to Golubeva et al.,
\cite{inc} also supports this conclusion.
Several studies \cite{phim} show the width of $\phi$ meson in a nucleus
increases significantly, where as the mass-shift for it is insignificant.

Another way to explore the $\rho$ dynamics in a normal nucleus could
be through the coherent $\rho^0$ production in the $(p,p^\prime)$ reaction.
The coherent meson production process had been used earlier to unveil many
interesting aspects in the nuclear dynamics.
For example, the pion and delta dynamics in a nucleus were investigated
sometime back through the coherent photo- and electro-production of pion
in the $\Delta$ excitation region \cite{erwe,pid,empi}. In this energy
region, the measurements had been done, in the recent past, on the coherent
$\pi^+$ production in the charge exchange reactions \cite{cpiex}. These
reactions offer valuable informations \cite{cpith}, which are complementary
to those obtained from the scattering of real pion on a nucleus
\cite{erwe,pid}.

For the coherent $\rho^0$ emission in the $(p,p^\prime)$ reaction, it is
visualized that the beam proton emits rho meson. This rho meson, of course,
is a virtual particle, since the four-momentum carried by it is constrained
by the dispersion relation obeyed by protons.
It is the nucleus which scatters this $\rho^0$ meson coherently and brings
it on-shell. Kinematically, this process is possible because the recoiling
nucleus, as a whole, can adjust the momentum to put the virtual rho meson
on its mass-shell.
For simplicity, we restrict the propagation of the coherent $\rho^0$ meson
to the forward direction only. This rho meson, after traveling a certain
distance, decays into $\pi^+$ and $\pi^-$ in the final state. Naturally, the
$\rho^0$ dynamics in this reaction  can be manifested from these $\pi^+\pi^-$
events. Therefore, the coherent rho meson production process can be thought
as a pertinent probe to acquire the knowledge about the $\rho$ nucleus
interaction.

The energy transfer $E_p-E_{p^\prime}$ distribution spectra have been
calculated for the $(p,p^\prime [\pi^+\pi^-])$ reaction on various nuclei.
According to the model presented here, the $\pi^+\pi^-$ in the final state
arise due to the decay of coherent $\rho^0$ meson, propagating through a
nucleus.
This calculation is based on Glauber approach for the nuclear effect. Where
the distorted wave functions for the continuum particles (i.e., $p,
~p^\prime, ~\pi^+ ~ \mbox{and} ~\pi^-$) and the $\rho^0$ propagator have
been described by the eikonal form. The potential required to generate
these quantities is constructed by using $``t\varrho"$
approximation.
In this study, the dependence of the cross section on the pion emission angle
as well as on the beam energy is presented. The cross sections for the rho
meson decaying inside and outside the nucleus are compared. Along with this,
the coherent and incoherent contributions to the cross section due to the
inside and outside decay for the rho meson are also studied.
The initial and final state interactions for this reaction are investigated.
In addition, the medium effect on the rho meson propagating through the
nucleus is explored.

We write the formalism in Sec.~2 for the coherent rho meson production in
the $(p,p^\prime)$ reaction and its decay into $\pi^+$ and $\pi^-$ in the
final state. The results and discussion for this reaction are illustrated
in Sec.~3. Conclusions obtained from this study have been presented in
Sec.~4.

\section {Formalism}
\label {for}

~~~~
The formalism for the $\pi^+\pi^-$ emission from the rho meson,
propagating through a nucleus, has been developed here. This rho meson is
produced coherently in the $(p,p^\prime)$ reaction on a nucleus. The
$T$-matrix $T_{fi}$ for this reaction consists of production,
propagation, and decay (into two pions) for the rho meson, i.e.,
\begin{eqnarray}
T_{fi} = \int \int d{\bf r} d{\bf r^\prime}
\chi^{(-)*} ({\bf k_{\pi^+}, r^\prime})
\chi^{(-)*} ({\bf k_{\pi^-}, r^\prime})
\Gamma_{\rho \pi \pi} S_\rho ( {\bf r^\prime - r} )
\Psi_{\rho^0} ({\bf r}).
\label{tmx}
\end {eqnarray}

The factor $ \Psi_{\rho^0} ({\bf r}) $ in the above equation represents the
$\rho^0$ production amplitude due to $(p,p^\prime)$ reaction. It is given
by
\begin{eqnarray}
\Psi_{\rho^0} ({\bf r}) =  \Pi_\rho (q_0, {\bf r}) G_\rho (q^2)
\chi^{(-)*} ({\bf k_{p^\prime}, r})
\Gamma_{\rho NN} \chi^{(+)} ({\bf k_p, r}),
\label{rpd}
\end {eqnarray}
where
$ \Pi_\rho (q_0, {\bf r}) [ = 2q_0V_{O\rho} ({\bf r}) ] $ denotes the
self-energy for the rho meson, arising due to $\rho$ nucleus
optical potential $ V_{O\rho} ({\bf r}) $. This potential describes the
elastic scattering of the virtual rho meson to its real state.
$q_0 (=E_p-E_{p^\prime})$ is the energy carried out by the virtual rho meson.
$ {\tilde G}_\rho (q^2) $ represents the propagator for the virtual rho
meson emitted at the $\rho^0pp^\prime$ vertex:
$ {\tilde G}_\rho (q^2) = - \frac{ 1 }{ m^2_\rho-q^2 } $. Here, $m_\rho$
($\sim 770$ MeV) and $ q (=k_p -k_{p^\prime}) $ are the mass and the four
momentum respectively for the rho meson. $\chi$s, given in Eq.~(\ref{rpd}),
denote the distorted wave functions for protons. In eikonal approximation,
they can be written as
\begin{eqnarray}
\chi^{(+)} ( {\bf k}_p, {\bf r} ) = e^{i {\bf k}_p . {\bf r} }
exp \left [ -\frac{i}{v_p}
\int^z_{-\infty} dz^\prime V_{Op} ({\bf b}, z^\prime) \right ],
\nonumber
\end{eqnarray}
and
\begin{equation}
\chi^{(-)*} ( {\bf k}_{p^\prime}, {\bf r} )
= e^{-i {\bf k}_{p^\prime} . {\bf r} }
exp \left [ -\frac{i}{v_{p^\prime}}
\int^\infty_z dz^\prime V_{O{p^\prime}} ({\bf b}, z^\prime)  \right ].
\label{pwfn}
\end{equation}
$\Gamma_{\rho NN}$, in Eq.~(\ref{rpd}), is the vertex function at
the $\rho^0pp^\prime$ vertex. It describes the emission of virtual $\rho^0$
at this vertex. The Lagrangian, describing the interaction at the $\rho NN$
vertex, is given by
\begin{equation}
{\cal L}_{\rho NN} = -g_VF(q^2) {\bar N}
[ \gamma^\mu - i\frac{ \kappa }{ 2m_N } \sigma^{\mu\nu} q_\nu ]
{\bf \tau \cdot \rho_\mu} N,
\label{lrnn}
\end {equation}
with $g_V=2.6$ and $\kappa=6.1$ \cite{erwe}. $F(q^2)$ represents the form
factor at the $\rho NN$  vertex:
$ F(q^2) = \frac{ \Lambda^2-m^2_\rho }{ \Lambda^2-q^2 } $,
with $ \Lambda=1.3 $ GeV/c \cite{mach}. The quantities $m_\rho$ and $q$ are
already defined below the Eq.~(\ref{rpd}).

The factor $ S_\rho ({\bf r^\prime - r}) $, appearing in Eq.~(\ref{tmx}),
describes the propagation for the real $\rho^0$ meson from the position
${\bf r}$ to another position ${\bf r^\prime}$. Which, in the present work,
has been restricted to the forward direction only. Therefore,
$ S_\rho ({\bf r^\prime - r}) $ can be expressed in the eikonal form as
\begin{eqnarray}
S_\rho ( {\bf r^\prime - r} )
= \delta ( {\bf b^\prime - b} ) \Theta (z^\prime - z)
e^{ i {\bf k_\rho . (r^\prime - r)} } D_{\bf k_\rho} ({\bf b}, z^\prime, z),
\label{rrp}
\end {eqnarray}
where $ D_{\bf k_\rho} ({\bf b}, z^\prime, z) $ describes the nuclear medium
effect on the $\rho$ propagation through a nucleus. It is given by
\begin{equation}
D_{\bf k_\rho} ({\bf b}, z^\prime, z) = - \frac{ i }{ 2k_\rho }
exp \left [  \frac{ i }{ 2k_\rho } \int_z^{z^\prime}
 dz^{\prime \prime} \{ {\tilde G}^{-1}_{0\rho} ( m ) -
    2 E_\rho V_{O\rho} ({\bf b}, z^{\prime \prime}) \} \right  ].
\label{drh}
\end{equation}
In this equation, $k_\rho$ is the momentum for the forward going rho meson.
$ V_{O\rho} ({\bf b}, z^{\prime \prime}) $, as mentioned earlier, represents
the optical potential for the rho meson.
$ {\tilde G}^{-1}_{0\rho} ( m ) $ represents the inverse of the free rho
meson (on-shell) propagator. It is given by
\begin{equation}
{\tilde G}^{-1}_{0\rho} ( m )
= m^2 - m^2_\rho +i m_\rho \Gamma_\rho ( m );
~~~
\Gamma_\rho ( m )
= \Gamma_\rho ( m_\rho ) \left ( \frac{ m_\rho }{ m } \right )
  \left [ \frac{ k ( m^2 ) }{ k ( m^2_\rho ) } \right ]^3,
\label{wdrh}
\end {equation}
with  $ m_\rho \approx 770 $ MeV and $ \Gamma_\rho ( m_\rho ) \sim 150 $
MeV, for $ \rho^0 \to \pi^+\pi^- $ in the free state. $k(m^2)$ is the
momentum for pion in the rest frame of decaying rho meson of mass $m$.

$\Gamma_{\rho\pi\pi}$, in Eq.~(\ref{tmx}), describes the vertex function
for the $\rho^0$ meson decaying into $\pi^+\pi^-$ in the final state. It is
governed by the Lagrangian:
\begin{equation}
{\cal L}_{\rho \pi \pi}
=  f_{\rho \pi \pi} {\bf \rho^\mu \cdot (\pi \times \partial_\mu \pi) },
\label{lrpi}
\end {equation}
with $ f_{\rho\pi\pi} \approx 6.1 $ \cite{erwe, rkb}.

The distorted wave functions $\chi$s, in Eq.~(\ref{tmx}), for pions in the
final state have been described by the eikonal form, i.e.,
\begin{eqnarray}
\chi^{(-)*} ({\bf k_{\pi^+},r^\prime})
\chi^{(-)*} ({\bf k_{\pi^-},r^\prime})
=
e^{ -i {\bf (k_{\pi^+}+k_{\pi^-}) . r^\prime} }
D_{\bf k_\pi} ({\bf b^\prime}, z^\prime).
\label{piwfn}
\end {eqnarray}
In this equation, $ D_{\bf k_\pi} ({\bf b^\prime}, z^\prime) $ represents
the distortion occurring due to the elastic scattering of pions by the
recoiling nucleus. It is given by
\begin{equation}
D_{\bf k_\pi} ({\bf b^\prime}, z^\prime)
= exp \left [ -i \int_{z^\prime}^\infty
dz^{\prime \prime} \left \{
  \frac{ V_{O\pi^+} ({\bf b^\prime}, z^{\prime \prime}) }{ v_{\pi^+} }
+ \frac{ V_{O\pi^-} ({\bf b^\prime}, z^{\prime \prime}) }{ v_{\pi^-} }
   \right \}  \right ],
\label{dpi}
\end{equation}
where $v_{\pi^\pm}$ is the velocity of $\pi^\pm$, and
$ V_{O\pi^\pm} ({\bf b^\prime}, z^\prime) $ denotes the optical potential
for the $\pi^\pm$ scattering on a nucleus.

The $T$-matrix $T_{fi}$, given in Eq.~(\ref{tmx}), can be factorized into
interaction and structure parts. The interaction part contains the
production and decay vertex functions
(i.e., $ \Gamma_{\rho NN} \Gamma_{\rho\pi\pi}$) for the rho meson. The
structure part carries the information about the medium effect on the rho
meson. The later, using Eqs.~(\ref{tmx}), (\ref{rrp}) and (\ref{piwfn}), can
be written as
\begin{equation}
F = \int d{\bf r} D_{\pi\rho} ({\bf k_\pi, k_\rho} ; {\bf b},z)
  e^{-i{\bf k_\rho . r}}  \Pi_\rho (q_0, {\bf r})
 {\tilde G}_\rho (q^2)
  \chi^{(-)*} ({\bf k_{p^\prime}, r}) \chi^{(+)} ({\bf k_p, r}),
\label{frh}
\end {equation}
with $ {\bf k_\rho = k_{\pi^+} + k_{\pi^-} } $. The symbol
$ D_{\pi\rho} ({\bf k_\pi, k_\rho} ; {\bf b},z) $ appearing in this equation
stands for
\begin{equation}
D_{\pi\rho} ({\bf k_\pi, k_\rho} ; {\bf b},z)
= \int^\infty_z dz^\prime  D_{\bf k_\pi} ({\bf b}, z^\prime)
      D_{\bf k_\rho} ({\bf b}, z^\prime, z),
\label{dpr}
\end {equation}
where $ D_{\bf k_\rho} ({\bf b}, z^\prime, z) $ and
$ D_{\bf k_\pi} ({\bf b}, z^\prime) $ are defined in Eqs. (\ref{drh}) and
(\ref{dpi}) respectively. The above equation describes the dynamics for
the pi and rho mesons in the nucleus.

The rho meson, due to its  short life time ($\sim 10^{-23}$ sec), can
decay inside as well as outside the nucleus, depending upon its velocity and
the size of the nucleus. Indeed, the physical quantity which determines this
decay probability is the effective decay length:
$ \lambda^* = v/\Gamma^* $. Where $v$ and $\Gamma^*$ describe the velocity
and the effective decay width respectively for the rho meson in a nucleus.
This
decay length $\lambda^*$ controls over the attenuation for the in-medium
$\rho$ propagation. While the rho meson propagates a distance
$ L({\bf b},z) [ = \sqrt { R^2-b^2 } - z] $ from the production point $z$ to
the surface of a nucleus of radius $R$, it gets attenuated by a factor
$ exp [ -L({b},z)/\lambda^* ] $.
To
investigate the inside and outside decay probabilities more transparently,
$ D_{\pi\rho} ({\bf k_\pi, k_\rho} ; {\bf b},z) $ in Eq.~(\ref{dpr}) is
splitted into two parts, i.e.,
\begin{equation}
D_{\pi\rho} ({\bf k_\pi, k_\rho} ; {\bf b},z)
=  D^{in}_{\pi\rho} ({\bf k_\pi, k_\rho} ; {\bf b},z)
 + D^{out}_{\pi\rho} ({\bf k_\pi, k_\rho} ; {\bf b},z).
\label{dio}
\end {equation}
In this equation, $ D^{in}_{\pi\rho} ({\bf k_\pi, k_\rho} ; {\bf b},z) $
and $ D^{out}_{\pi\rho} ({\bf k_\pi, k_\rho} ; {\bf b},z) $
correspond to the probability for the rho meson decaying inside and outside
the nucleus respectively. They are given by
\begin{equation}
D^{in}_{\pi\rho} ({\bf k_\pi, k_\rho} ; {\bf b},z)
= \int^{\sqrt{R^2-b^2}}_z dz^\prime  D_{\bf k_\pi} ({\bf b}, z^\prime)
      D_{\bf k_\rho} ({\bf b}, z^\prime, z),
\label{idpr}
\end {equation}
and
\begin{equation}
D^{out}_{\pi\rho} ({\bf k_\pi, k_\rho} ; {\bf b},z)
= \int^\infty_{\sqrt{R^2-b^2}} dz^\prime  D_{\bf k_\pi} ({\bf b}, z^\prime)
      D_{\bf k_\rho} ({\bf b}, z^\prime, z).
\label{odpr}
\end {equation}
In the later equation, i.e., in Eq.~(\ref{odpr}),
$ D_{\bf k_\pi} ({\bf b}, z^\prime) $ goes to unity, and
$ D_{\bf k_\rho} ({\bf b}, z^\prime, z) $ reduces to
$ -\frac{i}{2k_\rho} exp [ \frac{i}{2k_\rho} {\tilde G}^{-1}_{0\rho}
(m_\rho) (z^\prime-z) ]. $

For the $\pi^+\pi^-$ emission in the $(p,p^\prime)$ reaction on a nucleus,
the differential cross section,
$ \frac { d\sigma } { dE_{p^\prime} d\Omega_{p^\prime}
dE_{\pi^+} d\Omega_{\pi^+} d\Omega_{\pi^-} } $, for the energy transfer
transfer $E_p-E_{p^\prime}$ distribution can be written as
\begin{equation}
\frac { d\sigma } { dE_{p^\prime} d\Omega_{p^\prime}
dE_{\pi^+} d\Omega_{\pi^+} d\Omega_{\pi^-} }
= \frac{\pi^3}{(2\pi)^{11}}
  \frac{ m^2_p k_{p^\prime} k_{\pi^+} k_{\pi^-} }{ k_p } <|T_{fi}|^2>.
\label{dsg}
\end {equation}
The annular bracket around the $|T_{fi}|^2$ represents the average over the
spin orientations of the particles in the initial state and the summation
over the spin orientations of the particles in the final state.

\section {Results and Discussion}
\label {rlds}

~~~~
To calculate the $T$-matrix $T_{fi}$, given in Eq.~(\ref{tmx}), it is
necessary to evaluate the wave functions for protons in Eq.~(\ref{pwfn})
and pions in Eq.~(\ref{piwfn}), as well as the propagator for the rho meson
in Eq.~(\ref{rrp}). The important ingredient required to generate these
quantities is the optical potential for the respective particles. This
potential, in the present work, has been estimated using ``$t\varrho$''
approximation.  According to it, the optical potential
$ V_{OX} ({\bf r}) $ for a particle $X$ scattered off by a nucleus can be
expressed \cite{crp} as
\begin{equation}
V_{OX} ({\bf r}) = - \frac{v_X}{2} (i+\alpha_{XN}) \sigma_t^{XN}
                         \varrho ({\bf r}).
\label{opt}
\end{equation}
$\alpha_{XN}$ appearing in this equation denotes the ratio:
$ Re f_{XN}(0) / Im f_{XN}(0) $. Here, $f_{XN}(0)$ is the particle nucleon
$(X-N)$ scattering amplitude at the forward direction. $\sigma_t^{XN}$
represents the total scattering cross section. $\varrho ({\bf r})$ denotes
the spatial distribution for the nuclear density, usually approximated by
the charge density distribution for the nucleus.
The present work deals with $^{12}$C and $^{208}$Pb nuclei only. Therefore,
the form of $\varrho ({\bf r})$ for these nuclei, as extracted from the
electron scattering data \cite{at}, is given by
\begin{eqnarray}
&^{12}\mbox{C} ~~ :&
\varrho ({\bf r}) = \varrho_0 [ 1 + w(r/c)^2 ] e^{ -(r/c)^2 };
~~ w=1.247,  ~ c=1.649 ~ \mbox{fm}.  ~~~~~ ~~~~~ ~~~~~      \\
\label{den1}
&^{208}\mbox{Pb}:&
\varrho ({\bf r})
= \varrho_0 \frac{1+w(r/c)^2}{1+exp(\frac{r^2-c^2}{a^2})};
~ w=0.3379,  c=6.3032 ~\mbox{fm}, a=2.8882 ~\mbox{fm}.  ~~~~~ ~~
\label{den2}
\end{eqnarray}
These densities are normalized to the mass numbers of the corresponding
nuclei.

To evaluate the proton nucleus optical potential $ V_{Op} ({\bf r}) $
in Eq.~(\ref{opt}), the energy dependent measured values for the proton
nucleon scattering parameters, i.e., $ \alpha_{pN} $ and $\sigma_t^{pN}$,
\cite{scpn, pdg} have been used.
Similarly, $\alpha_{\pi^\pm N}$ and $\sigma_t^{\pi^\pm N}$ are used as
input quantities in Eq.~(\ref{opt}) to generate the $\pi^\pm$ nucleus
optical potential $ V_{O\pi^\pm} ({\bf r}) $. The data for
$\sigma_t^{\pi^\pm p}$ exist for a wide range of energy \cite{pdg}. The
measured values for $\alpha_{\pi^\pm p}$ are also available in
Ref.~\cite{pdg} at higher energies only. Therefore, $\alpha_{\pi^\pm p}$ has
been generated at lower energies by using SAID program \cite{said}.
For $\pi^\pm$ neutron scattering parameters, they are approximated as:
$ \alpha_{\pi^\pm n} \approx \alpha_{\pi^\mp p} $ and
$ \sigma_t^{\pi^\pm n} \approx \sigma_t^{\pi^\pm d } -
\sigma_t^{\pi^\pm p} $. The data for the pion deuteron total scattering
cross section $\sigma_t^{\pi^\pm d}$ are also available in the
Ref.~\cite{pdg}.

The rho meson, being an unstable particle, can not exist as beam. Therefore,
$\alpha_{\rho^0 N}$ and $\sigma_t^{\rho^0 N}$ [which are required in
Eq.~(\ref{opt}) to estimate the $\rho^0$ nucleus optical potential
$V_{O\rho} ({\bf r})$] can not be measured directly. On the other hand, they
can be extracted from the $\rho$ production data available for various
elementary reactions. In fact, several authors have done this exercise
recently for a wide range of $\rho$ energy.
In one of these calculations, using Vector Dominance Model (VDM) Kondratyuk
et al., \cite{kscge} have extracted the $\rho$ nucleon scattering amplitude
$f_{\rho N}$  from the rho meson photoproduction data at higher energies
$(E_\rho \ge 2 ~ \mbox {GeV})$.
At lower energies $(E_\rho \le 2 ~ \mbox {GeV})$ they have calculated
$f_{\rho N}$ using Resonance Model (RM), since the rho meson couples strongly
to a number of resonances in this energy region.
There exists another calculation, where Lutz et al., using couple channel
approach \cite{lwf}, have calculated $f_{\rho N}$ in the region of threshold
$\rho$ production. In this energy region, their calculations reproduce the
measured cross sections for various reactions, such as
$\pi N \to \rho N, ~ \gamma N \to \rho N$, very well.

As mentioned earlier, we calculate the energy transfer $E_p-E_{p^\prime}$
distribution spectra for the $\pi^+\pi^-$ emission, which arise due to the
decay of the forward going $\rho^0$ meson in the nucleus. This rho meson is
assumed to produce coherently in the $(p,p^\prime)$ reaction on a nucleus.
The coherent meson production process assures that the recoil nucleus remains in the
same state as of the target nucleus. The forward propagation of the $\rho^0$
meson, as considered here for simplicity, imposes a constrain on the pion
emission angle: $ \theta_{\pi^+} = -\theta_{\pi^-} $. In these calculations,
the $p^\prime$ emission angle is taken fixed at $1^0$.

For the above reaction on $^{12}$C nucleus, the sensitivity of the cross
section on the pion emission angle is presented in Fig.~1 at 2.5 GeV beam
energy. The cross sections have been calculated for various $\pi^+$
emission angles $\theta_{\pi^+}$ taken sequentially from $5^0$ to $35^0$. In
this figure, we show the calculated results only for $\theta_{\pi^+}$ taken
equal to $ 20^0, ~25^0 ~\mbox{and} ~30^0 $.
This figure shows that the peak position gets shifted  towards the lower
value of $E_p-E_{p^\prime}$ with the increase in the $\pi^+$ emission angle.
It also shows the cross section is maximum for the $\pi^+$ emission angle
taken equal to $25^0$.
For this angle, the cross section at the peak is 82.5 $\mu$b/(GeV$^2$sr$^3$),
which appears at the energy transfer $E_p-E_{p^\prime}$ equal to 1.64 GeV.
The peak cross section for $\theta_{\pi^+}$ taken equal to $20^0$ is 52.35
$\mu$b/(GeV$^2$sr$^3$), appearing at $E_p-E_{p^\prime}$ equal to 1.74 GeV.
Where as, the cross section at the peak for $\theta_{\pi^+}$ taken at $30^0$
is about 48 $\mu$b/(GeV$^2$sr$^3$), and it appears at $E_p-E_{p^\prime}$
equal to 1.43 GeV.

In nuclear reactions, it is often seen that the cross section strongly
depends on the beam energy. To show this dependence, the energy transfer
$E_p-E_{p^\prime}$ distribution spectra have been calculated for the above
reaction at beam energies taken equal to 2.5, 3 and 3.5 GeV. The $\pi^+$
emission angle $\theta_{\pi^+}$ is taken fixed at $25^0$, since this
settings, as shown in Fig.~1, gives the maximum cross section at 2.5 GeV
beam energy.
The calculated $E_p-E_{p^\prime}$ distribution spectra are presented in
Fig.~2, which shows the cross section increases with the increase in beam
energy. The cross section at the peak is enhanced to 0.75 mb/(GeV$^2$sr$^3$)
at 3.5 GeV from 82.5 $\mu$b/(GeV$^2$sr$^3$) at 2.5 GeV. The peak position at
3.5 GeV beam energy appears at $E_p-E_{p^\prime}$ equal to 1.74 GeV, where as
it appears at $E_p-E_{p^\prime} = 1.64 $ GeV for the beam energy taken equal
to 2.5 GeV. At 3 GeV, the calculated cross section at the peak is 0.4
mb/(GeV$^2$sr$^3$), and it appears at $E_p-E_{p^\prime}$ equal to 1.7 GeV.

An unstable particle, while propagating through a nucleus, can decay inside
as well as outside the nucleus, depending upon its velocity and the size of
the nucleus. To explore it for $\rho^0$ meson propagating through the
$^{12}$C nucleus, the cross sections for the inside and outside decay
probabilities have been compared in Fig.~3 at 3.5 GeV beam energy. The $\pi^+$
emission angle is taken equal to $20^0$, since it gives the maximum cross
section (not shown) at 3.5 GeV beam energy.
The peak cross section for the rho meson decaying inside of this nucleus
is 0.12 mb/(GeV$^2$sr$^3$), appearing at $E_p-E_{p^\prime}$ equal to 1.98
GeV (dotted line). Where as it is equal to $\sim 1.46$ mb/(GeV$^2$sr$^3$)
at $E_p-E_{p^\prime} = 2.12 $ GeV for $\rho^0$ decaying outside the same
nucleus (dashed line). Therefore, the outside decay cross section for this
reaction dominates by a factor about 12 over the inside decay cross
section.
This figure also compares the coherent and incoherent contributions to the
cross section, coming from the inside and outside decay amplitudes for
$\rho^0$ traveling through the $^{12}$C nucleus. As illustrated in this
figure, the peak cross section due to incoherent contribution is about 1.57
mb/(GeV$^2$sr$^3$) at $ E_p-E_{p^\prime} = 2.12 $ GeV (dash-dot line). On
the other hand, it is $\sim 2.17$ mb/(GeV$^2$sr$^3$) due to coherent
contribution, appearing at $E_p-E_{p^\prime}$ equal to 2.08 GeV (solid line).
Therefore, the coherent contribution supercedes the incoherent contribution
by a factor about 1.4.

The inside decay probability for the rho meson should increase while it
propagates through a larger nucleus. To show this, the cross sections for
$\rho^0$ meson decaying inside as well as outside the $^{208}$Pb nucleus have been
calculated at 3.5 GeV beam energy. The $\pi^+$ emission angle is taken at
$15^0$ in this case, since it gives maximum cross section for the Pb target.
The
calculated results are presented in Fig.~4. It is noticeable in this figure
that the peak cross sections for the inside and outside decay probabilities
of the $\rho^0$ meson, traveling through the Pb nucleus, are comparable to
each other [$\sim$ 65 $\mu$b/(GeV$^2$sr$^3$)], unlike the case happening for
$^{12}$C nucleus (see in Fig.~3).
Another remarkable aspect appearing in this figure is the enhancement in the
cross section due to the interference between the inside and outside decay
amplitudes for the $\rho^0$ meson. The peak cross section due to incoherent
contribution is 0.11 mb/(GeV$^2$sr$^3$) at $E_p-E_{p^\prime}$ equal to 2.2
GeV (dash-dot). Where as, it is equal to 0.22 mb/(GeV$^2$sr$^3$) due to
coherent contribution, appearing at $ E_p-E_{p^\prime} = 2.19 $ GeV (solid
line). Therefore, the interference effect enhances the cross section over
the incoherent contribution by a factor of 2 for the Pb nucleus, which is
significantly larger than that for the C nucleus.

It is always very interesting to investigate the initial and final state
interactions in the nuclear reactions. The effect of these interactions on
the $\pi^+\pi^-$ emission in the $(p,p^\prime)$ reaction on $^{12}$C nucleus
is shown in Fig.~5 at 3.5 GeV beam energy. This figure compares the
calculated distorted wave results for the continuum particles (i.e., $ p,
~p^\prime, ~\pi^+ ~\mbox{and} ~\pi^- $) with the plane waves results. In
addition, the nuclear effect on the $\rho$ propagation through the $^{12}$C
nucleus is also presented.
The plane wave result (which also includes the free $\rho^0$ propagator)
shows the cross section is 25.54 mb/(GeV$^2$sr$^3$) at the peak, appearing
at $E_p-E_{p^\prime}$ equal to 2.02 GeV (dash-dot-dot line).
The distortion due to projectile proton $p$ brings down the peak cross
section to 7.02 mb/(GeV$^2$sr$^3$) at $ E_p-E_{p^\prime} = 2.04 $ GeV
(dash-dot line). Therefore, the initial state interaction reduces cross
section by a factor about 3.64 at the peak, and it shifts the peak position
by 20 MeV towards the higher side in the $E_p-E_{p^\prime}$ distribution
spectrum.
The incorporation of the distortion due to ejectile proton $p^\prime$ further
attenuates the peak cross section to 2.6 mb/(GeV$^2$sr$^3$) at
$ E_p-E_{p^\prime} = 2.05 $ GeV (dashed line). Therefore, the projectile and
ejectile (i.e., $p$ and $p^\prime$) distortions together reduce the plane
wave cross section by a factor about 9.82 at the peak, and they shift the
peak position by 30 MeV towards the higher value of $E_p-E_{p^\prime}$.
The
final state interactions due to $\pi^+\pi^-$ together shift the peak
position to $ E_p-E_{p^\prime} = 2.07 $ GeV, leaving the magnitude of the
peak cross section [$\sim 2.38$ mb/(GeV$^2$sr$^3$)] almost unaffected
(dotted line).
Therefore, the overall final state interactions reduce the cross section
by a factor about 2.95 at the peak, and they shift the peak position by 30
MeV towards the higher side in the $E_p-E_{p^\prime}$ distribution spectrum.
The initial and final state interactions together, as shown in this figure,
bring down the plane wave cross section by a factor $\sim 10.7$ at the peak,
and they also shift the peak position by 50 MeV towards the higher value in
the $E_p-E_{p^\prime}$ distribution spectrum.
The $\rho^0$ nucleus interaction (which appears in the rho meson propagator)
is found not much sensitive to the cross section. It attenuates the peak
cross section only by about $8.8\%$, i.e., from 2.38 mb/(GeV$^2$sr$^3$) to
2.17 mb/(GeV$^2$sr$^3$), and it shifts the peak position by 10 MeV to
$ E_p-E_{p^\prime} = 2.08 $ GeV (solid line).

The initial and final state interactions could be very intensive for those
nuclear reactions, which involve heavier nuclei. To illustrate it,
the calculated plane and distorted wave results for the Pb nucleus are
presented in Fig.~6.
As exhibited in this figure, the plane wave results (which also
incorporates the free $\rho^0$ propagator) show oscillations in the energy
transfer $E_p-E_{p^\prime}$ spectrum (dash-dot-dot line). It arises due to
the oscillatory nature of the form factor for the Pb nucleus.
The
initial state interaction reduces this oscillation drastically, showing a
prominent peak at 2.01 GeV in the energy transfer $E_p-E_{p^\prime}$
spectrum (dash-dot line). The cross section at this peak is about 0.74
(mb/GeV$^2$sr$^3$). Therefore, the initial state interaction alone brings
down the cross section at the main peak by a factor about 179, and it shifts
the peak position by 50 MeV towards the lower value of $E_p-E_{p^\prime}$.
The
incorporation of $p^\prime$ distortion reduces the peak cross section
further to 0.35 mb/(GeV$^2$sr$^3$), at $E_p-E_{p^\prime}$ equal to 2.07 GeV
(dashed line). Therefore, the projectile and ejectile (i.e., $p$ and
$p^\prime$)  distortions together attenuate the cross section at the main
peak by a factor about 379, and it shifts the peak position by 10 MeV only
towards the higher side in the $E_p-E_{p^\prime}$ distribution spectrum.
The
final state interactions due to $\pi^+\pi^-$ together flatten the peak
considerably, and they further reduce the peak cross section to $\sim 0.25$
mb/(GeV$^2$sr$^3$) (dotted line). Therefore, the final state interactions
altogether bring down the cross section by a factor about 3.
In
this case, the $\rho$ nucleus interaction reduces the cross section $12\%$,
i.e., from 0.25 mb/(GeV$^2$sr$^3$) to 0.22 mb/(GeV$^2$sr$^3$), leaving the
shape for the spectrum unchanged (solid line).

\section {Conclusion}
\label {cncl}
~~~~
This study shows that the cross section for the $\pi^+\pi^-$ emission from
the forward going coherent $\rho^0$ meson, produced in the $(p,p^\prime)$
reaction, reaches its maximum value for a certain direction of the pion
emission. The cross section strongly depends on the beam energy.
The probability for the rho meson decaying inside the nucleus is less for
lighter nucleus. It increases with the size of the nucleus. However, the
interference between the inside and outside decay amplitudes for the rho
meson is very important in evaluating the cross section. The initial and
final state interactions significantly dampen the plane wave cross section,
which is drastic for the heavier nucleus.

\section {Acknowledgement}
\label {ackn}
~~~~
I gratefully acknowledge Dr. A.K. Mohanty for his encouragement.

{\bf Figure Captions}
\begin{enumerate}

\item
The sensitivity of the energy transfer $E_p-E_{p^\prime}$ distribution
spectrum on the pion emission angle is shown in this figure. The peak cross
section is maximum for the $\pi^+$ emission angle taken equal to $25^0$. The
peak position gets shifted towards the lower value of $E_p-E_{p^\prime}$
with the increase in the pion emission angle.

\item
The dependence of the energy transfer $E_p-E_{p^\prime}$ distribution
spectrum on the beam energy is presented here. With the increase in the beam
energy, the cross section increases and the peak position moves towards the
higher value of $E_p-E_{p^\prime}$. The $\pi^+$ emission angle is taken fixed
at $25^0$ (see text).

\item
The cross sections for the rho meson decaying inside and outside the nucleus
are compared for $^{12}$C nucleus at 3.5 GeV beam energy. The outside decay
cross section superceds for this nucleus. The interference between the
inside and outside decay amplitudes enhances the cross section. Various
lines are explained in the text.

\item
The inside and outside decay cross sections are shown for the rho meson
propagating through the $^{208}$Pb nucleus. The inside decay cross section
is substantially increased for the heavier nucleus. In this case, the
interference due to the $\rho^0$ inside and outside decay amplitudes is more
stronger than that for the $^{12}$C nucleus.

\item
The effect of the initial and final state interactions on the cross section,
for the $(p,p^\prime[\pi^+\pi^-])$ reaction on the $^{12}$C nucleus, is
illustrated here. The beam energy is taken equal to 3.5 GeV. The nuclear
effect on the $\rho$ propagation is also presented. Various lines have been
described in the text.

\item
The initial and final state interactions arising due to $^{208}$Pb nucleus
are exhibited in this figure. Compare to $^{12}$C nucleus, these interactions
are more intense for the Pb nucleus. Various lines are defined in the text.

\end{enumerate}

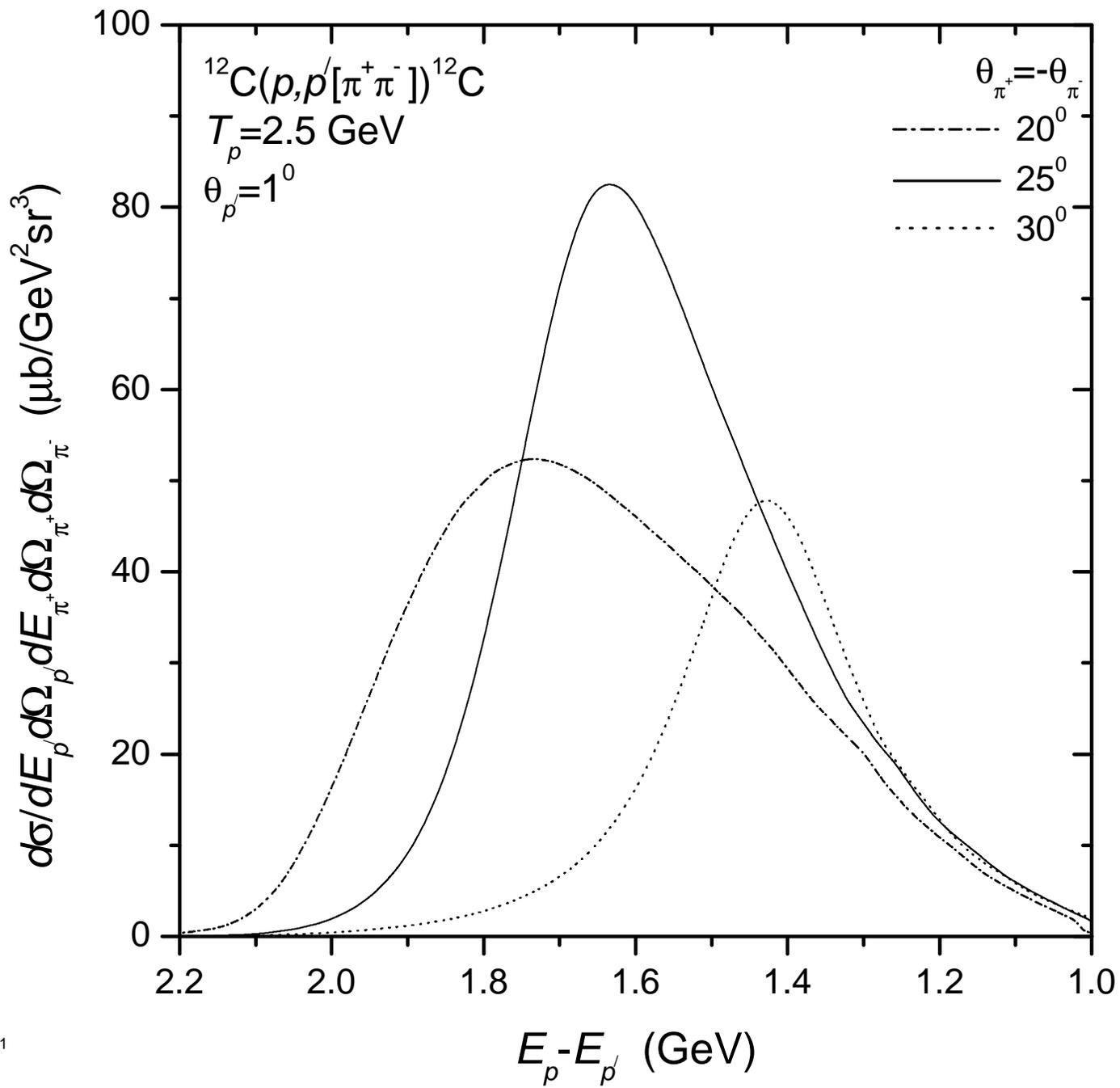
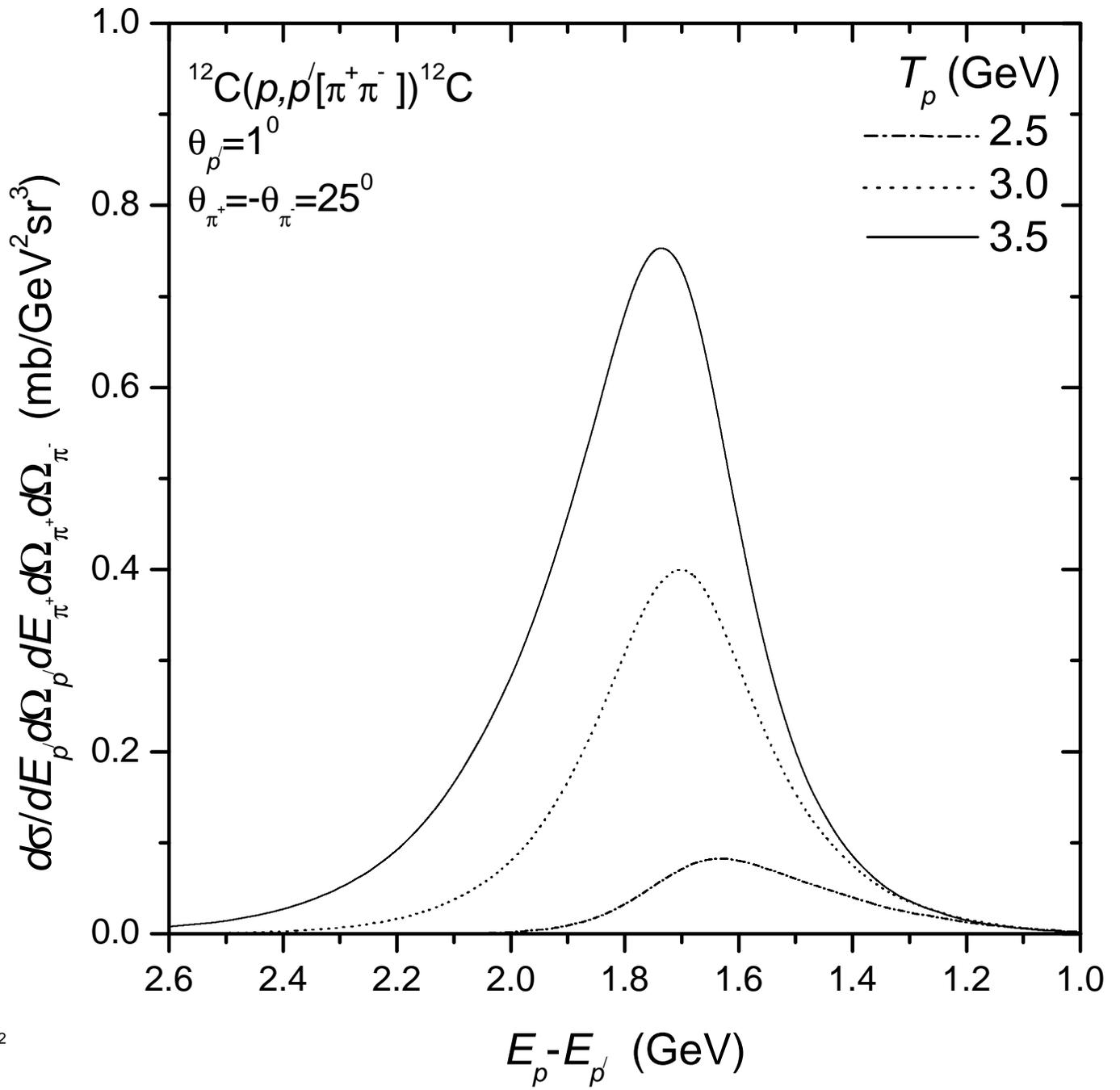
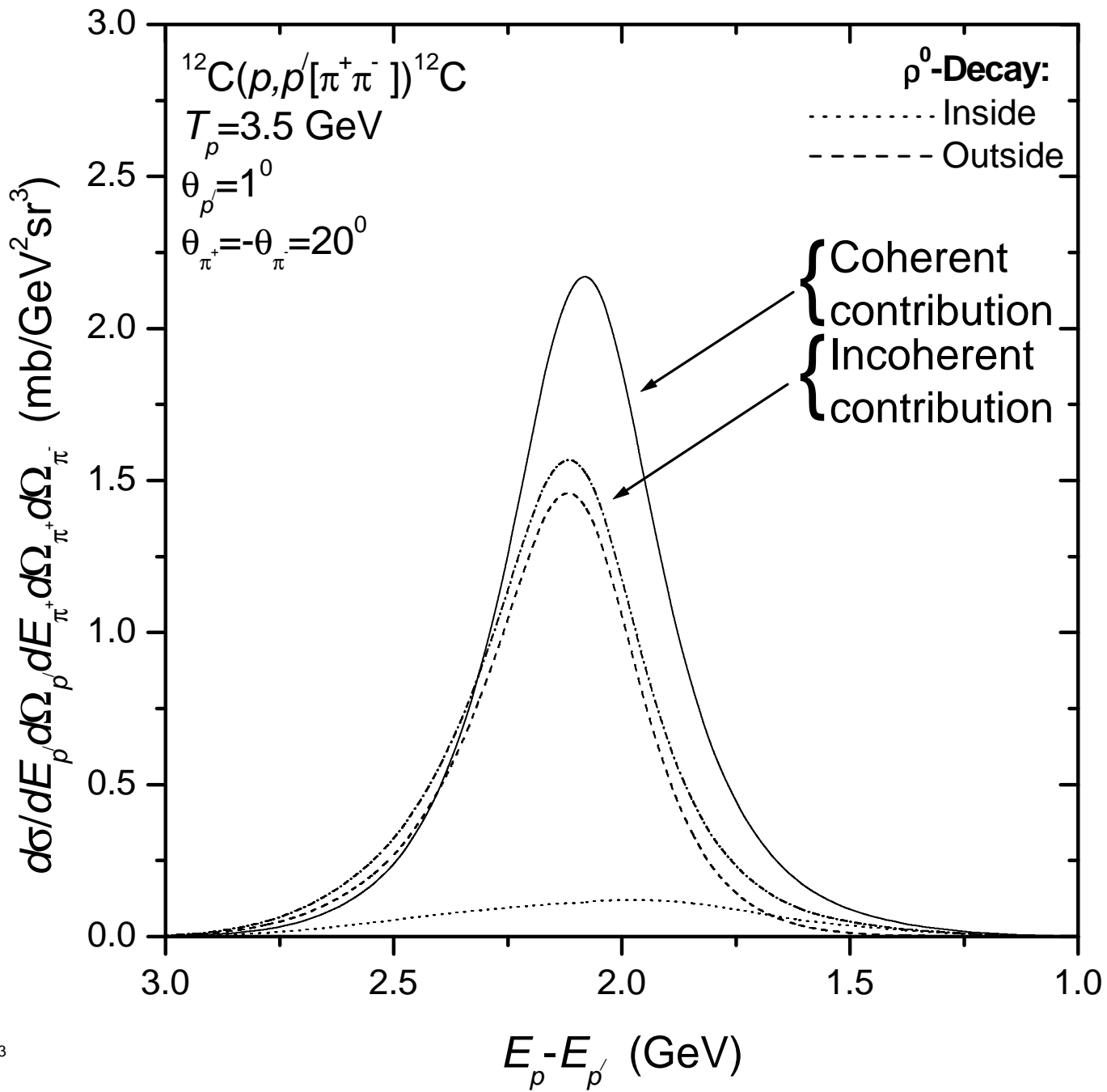
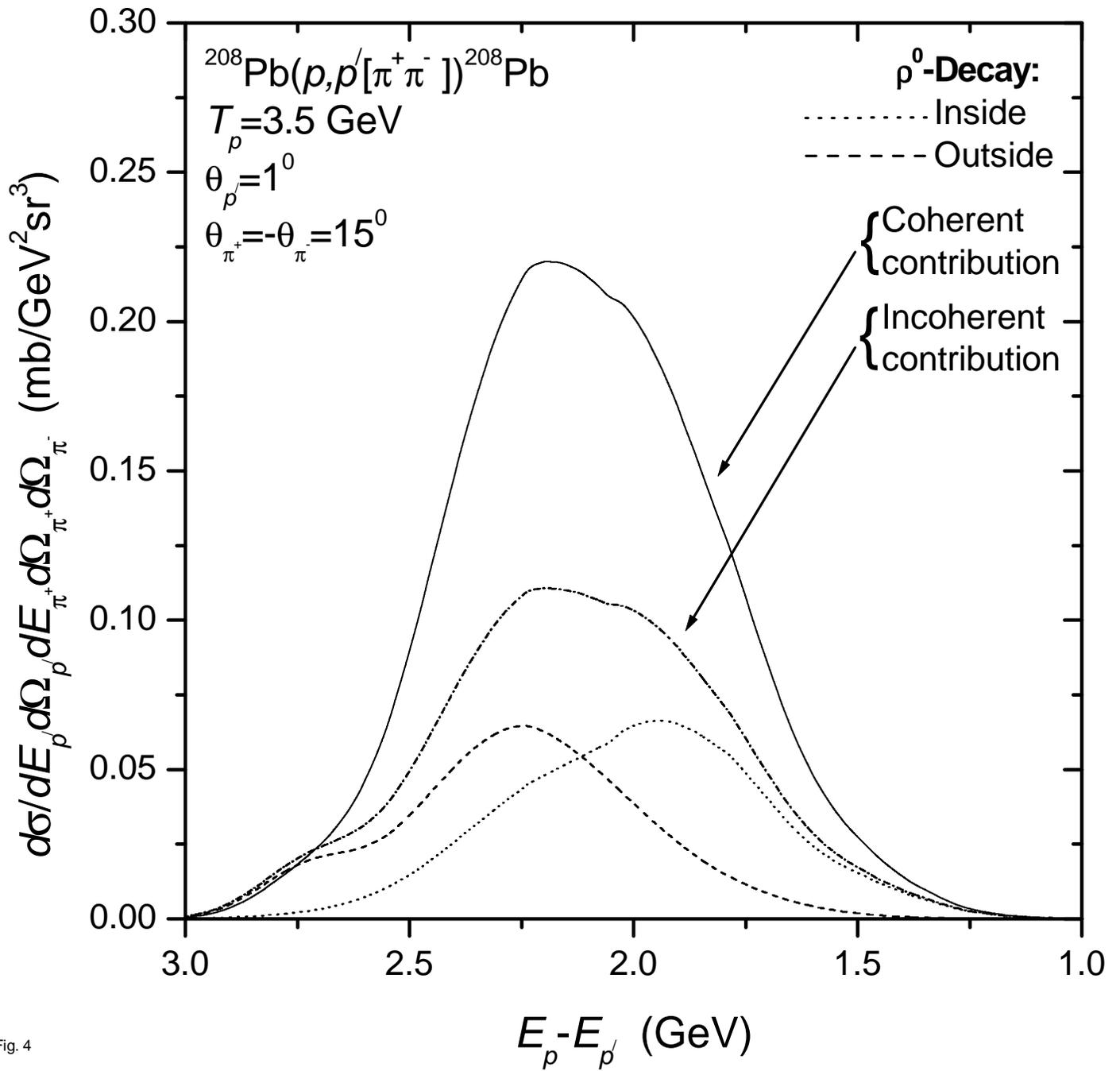
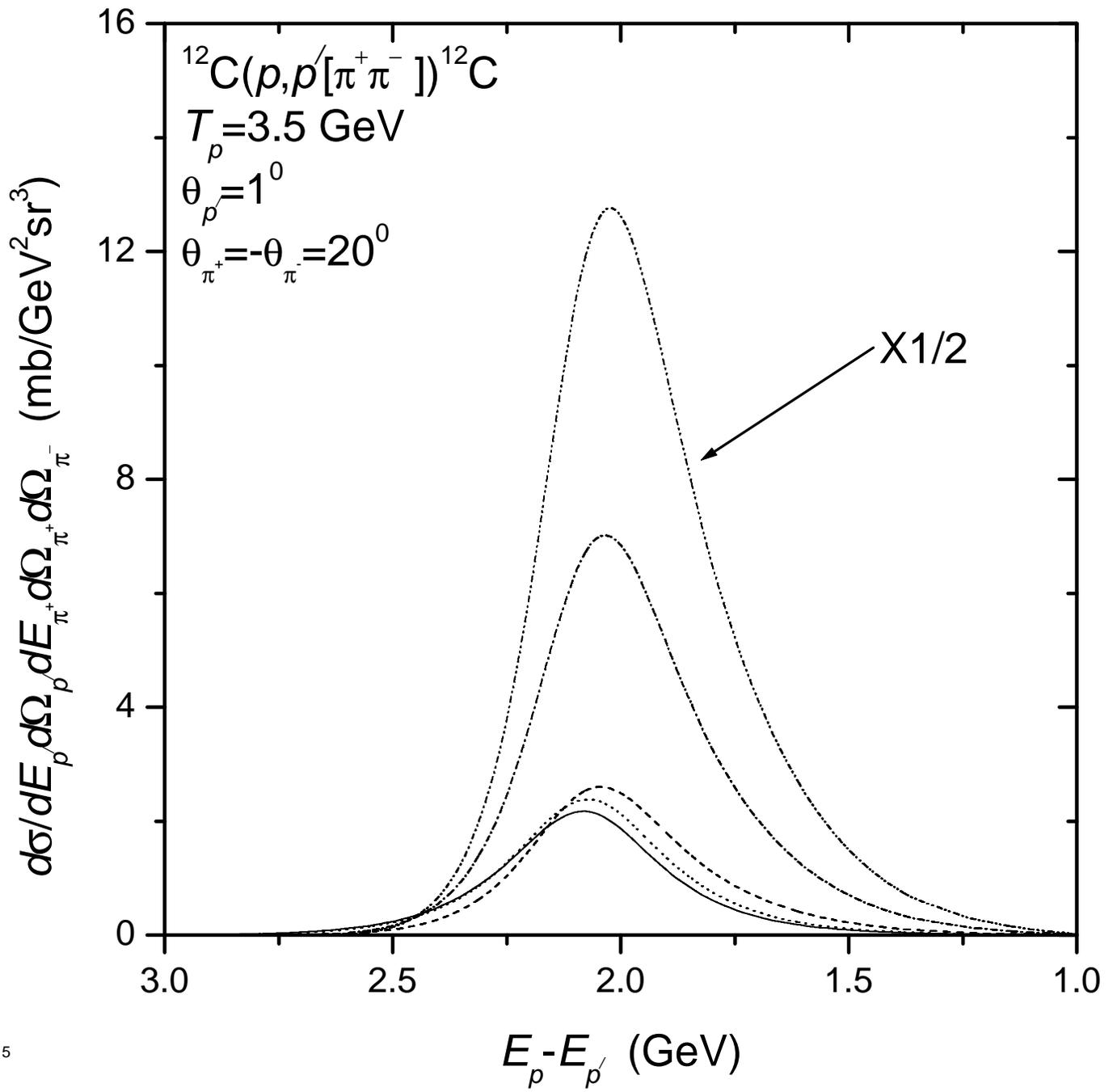
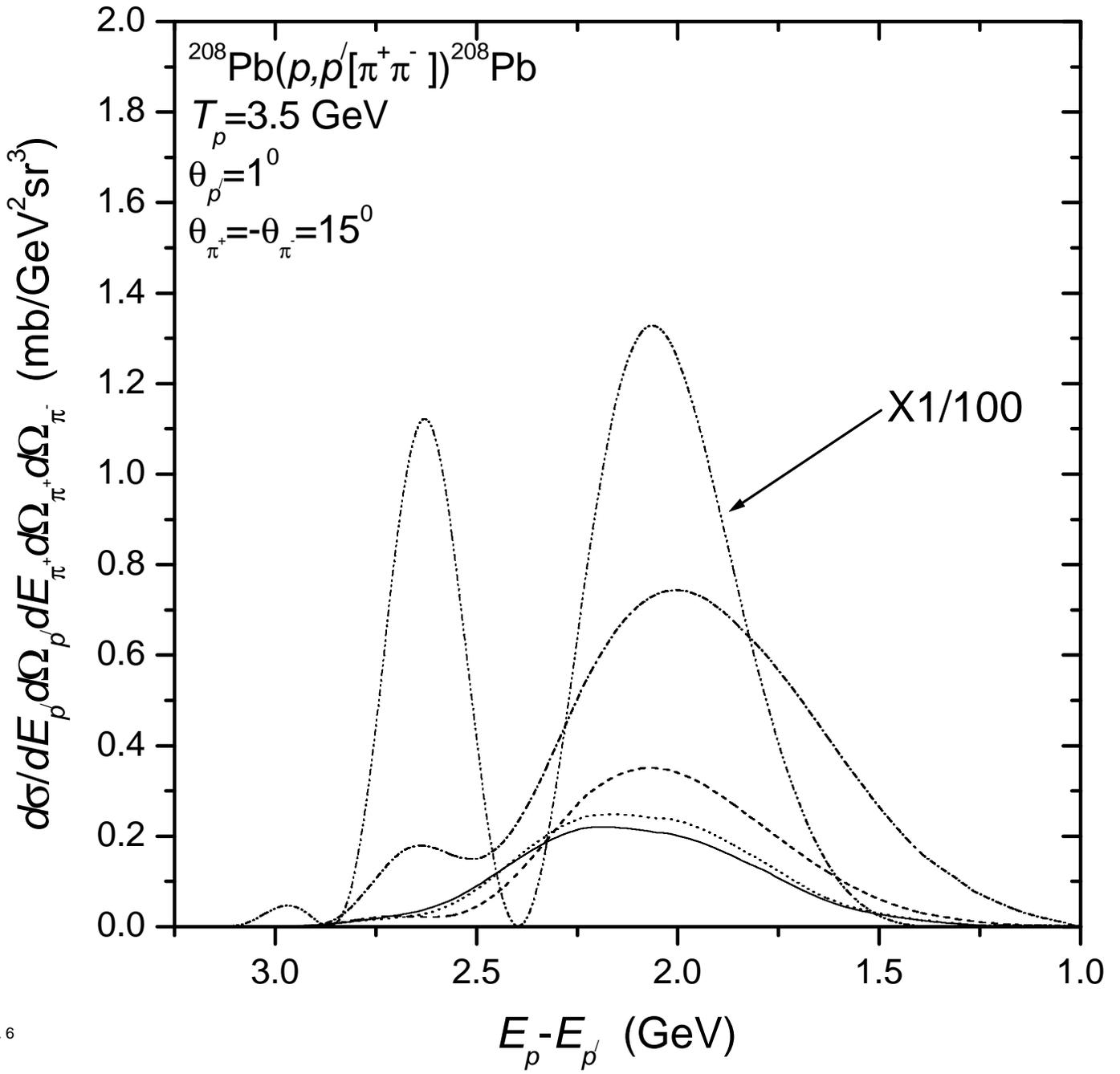
\end{document}